\documentclass[final,1p,times]{elsarticle}

\usepackage{graphicx}
\usepackage{amssymb}
\usepackage{color}

\date{}

\begin{document}

\begin{frontmatter}

\title{Studies of quasiclassical approach applicability to true three-body
decays}

\author[label1]{O.M.\ Sukhareva\footnote{o.m.sukhareva@gmail.com}$^,$}
\author[label2,label3,label4]{L.V.\ Grigorenko}
\author[label5,label6]{D.A.\ Kostyleva}
\author[label7]{M.V.\ Zhukov}

\address[label1]{Omsk State Technical University, 644050 Omsk, Russia}
\address[label2]{Flerov Laboratory of Nuclear Reactions, JINR, 141980 Dubna,
Russia}
\address[label3]{National Research Nuclear University ``MEPhI'', Kashirskoye
shosse 31, 115409 Moscow, Russia}
\address[label4]{National Research Centre ``Kurchatov Institute'', Kurchatov
sq.\ 1, 123182 Moscow, Russia}
\address[label5]{II. Physikalisches Institut, Justus-Liebig-Universit\"at, 35392
Giessen, Germany}
\address[label6]{GSI Helmholtzzentrum  f\"{u}r Schwerionenforschung GmbH, 64291
Darmstadt, Germany}
\address[label7]{Department of Physics, Chalmers University of Technology,
S-41296 G\"{o}teborg, Sweden}

\begin{abstract}
Within the hyperspherical harmonics approach the three-body problem is reduced
to a motion of one effective particle in a ``strongly deformed'' field, which is
described in coupled-channel formalism. This method is especially suited to
studies of phenomena characterized by genuine three-body dynamics, e.g.\
Borromean haloes and true three-body decays. The reduction of the hyperspherical
equations set to a single-channel Schr\"odinger equation provides the basis for
the use of the standard
quasiclassical expression for calculations of widths for true three-body decays.
We demonstrate that the quasiclassical approach by itself is quite precise in
application to typical profiles of the three-body effective potentials. However,
the reduction to single-channel formalism leads to significant overestimation of
the two-proton width $\Gamma_{2p}$. This is demonstrated by the example of the
$^{17}$Ne first excited $3/2^-$ state decay, questioning, however, the
applicability of such an approximation in general.
\end{abstract}


\end{frontmatter}


\section{Introduction}


Conventional methods of width determination for resonant states, such as elastic
phase shift energy dependence, or via S-matrix pole position in the complex
energy plane could be technically complicated for very small widths $\Gamma \ll
E$. Therefore, studies of radioactive decays require specific methods for the
decay width determination. Among them are ``natural'' width definition via wave
function (WF) with pure outgoing asymptotics \cite{Pfutzner:2012},
``Kadmensky-type'' integral formulas (IF) \cite{Grigorenko:2007},  and
quasiclassical (QC) approach of Gamow type \cite{Garrido:2004b}.

The use of a quasi-classical approach of Gamow type for the decay width
evaluation
\[
\Gamma \sim \exp \left[- 2 \int\nolimits_{r_2}^{r_3} p(r) \, dr \right]\,,
\]
requires the reduction of the few-body problem to a single-channel formalism of
some
form, where Gamow integral over the sub-barrier trajectory $\{r_2,r_3\}$ can be
defined. Here both the validity of the few-body problem reduction and the
applicability of the quasiclassical approximation for barriers of specific for
few-body physics shapes can be questioned.

The formalism of the Gamow type has been repeatedly used in recent years for the
determination of three-body decay widths
\cite{Garrido:2004b,Garrido:2005,Garrido:2005a,Garrido:2008,Garrido:2010,%
Garrido:2011,Hove:2016,Hove:2017}. In this work, we examine the validity of the
Gamow-type approximation by the example of the width of the first excited
$3/2^-$
state of $^{17}$Ne \cite{Garrido:2004b,Garrido:2008}. This state is known
to decay via the so-called ``true'' two-proton decay mechanism
\cite{Grigorenko:2000b,Pfutzner:2012}. There are several topics of interest
about this state discussed below in Section \ref{sec:motiv}. There is also a
certain story of theoretical controversy concerning width calculations for this
state \cite{Grigorenko:2003,Garrido:2004b,Grigorenko:2007,Garrido:2008}, see
Section \ref{sec:history}.


\subsection{Motivation for $^{17}$Ne $(3/2^-)$ decay studies}
\label{sec:motiv}


The $^{17}$Ne nucleus is a kind of ``test bench'' case for several interesting
concepts of nuclear structure and dynamics:

\noindent (i) The $^{17}$Ne $1/2^-$ ground state (g.s.) is not strongly bound.
The lowest-energy threshold is $2p$ one with $S_{2p}=0.933$ MeV. This is a
Borromean system since its core+$p$ subsystem, the $^{16}$F isotope, is
particle-unbound with $S_{p}=-0.535$ MeV. Because of its low binding, the
$^{17}$Ne has been considered as a candidate to possess $2p$ halo
\cite{Zhukov:1995c,Grigorenko:2003,Grigorenko:2005}. The question remains
open, see e.g.\ the discussions in Refs.\ \cite{Parfenova:2018,Wamers:2018}.

\noindent (ii) The $3/2^-$ first excited state of $^{17}$Ne is only slightly
unbound relative to $2p$ threshold with $Q_{2p}= 0.355$ MeV. Because none of the
$^{16}$F states is accessible for sequential proton emission, the decay mode of
this nucleus is ``true'' $2p$ decay \cite{Grigorenko:2000b,Pfutzner:2012}.
Because of the small $Q_{2p}$ the
radioactivity lifetime scale is expected for the $2p$ decay branch.
This possible decay mode is quite rare in the light nuclei and also the opportunity
to study such emission from an excited state is unique so far. Theoretical
calculations of this width have produced considerable controversy
\cite{Grigorenko:2003,Garrido:2004b,Grigorenko:2007,Garrido:2008} which we are
going to further discuss in this work.

\noindent (iii) There is a topic of interest from the nuclear astrophysics side,
as $^{15}$O is the rp-process ``waiting point''. The reaction $^{15}$O+$p$+$p
\rightarrow ^{17}$Ne+$\gamma$ provides a ``bypass'' of the $^{15}$O waiting
point together with the more ``conventional'' $^{15}$O+$\alpha \rightarrow
^{19}$Ne+$\gamma$ reaction \cite{Gorres:1995}.
In astrophysical conditions, the $^{15}$O+$p$+$p \rightarrow
^{17}$Ne+$\gamma$ reaction has two major reaction
mechanisms \emph{resonant}  and \emph{nonresonant}.

 The \emph{nonresonant}
contribution at temperatures of astrophysical interest is mainly defined by the
low-energy behavior of the E1 electromagnetic strength function. The latter has
the character of ``soft dipole mode'', closely connected with the halo
characteristics of $^{17}$Ne g.s.\ WF
\cite{Grigorenko:2006,Casal:2016,Parfenova:2018}. The \emph{resonant}
contribution to the $2p$ radiative capture on $^{15}$O is practically entirely
defined by the $2p$ width of the $3/2^-$ first excited state of $^{17}$Ne
\cite{Grigorenko:2005a,Sharov:2017}.


\subsection{History of the question}
\label{sec:history}


For the first time, the particle decay modes of $^{17}$Ne were experimentally
studied in paper \cite{Chromik:1997} with the first (as it was understood later,
erroneous) ideas about a possible observation of $2p$ emission form the $3/2^-$
state. These ideas were based on the $3/2^-$ state lifetime estimates obtained
in the diproton model: $\Gamma_{2p}=1.8 \times 10^{-12}$ MeV.

In 2000 the first quantum-mechanical theory of $2p$ radioactivity was developed
\cite{Grigorenko:2000b} in the framework of the three-body core+$p$+$p$ model
treated by the hyperspherical harmonics (HH) formalism. It was understood that
the
diproton model is giving the upper limit estimates for $2p$ widths, rather then
realistic results. The three-body model provided much smaller $\Gamma_{2p}=1.4
\times 10^{-15}$ MeV for $^{17}$Ne $3/2^-$ state. It became clear that there was
no chance to get evidence for $2p$ emission from $3/2^-$ in paper
\cite{Chromik:1997}.

The improved experiment of \cite{Chromik:2002} established the limit for the
ratio
of $2p$ and $\gamma$ widths for the $^{17}$Ne $3/2^-$ state $\Gamma_{2p} /
\Gamma_{\gamma} <7.7 \times 10^{-3}$. This ratio derivation was based on the
gamma width value $ \Gamma_{\gamma} = 5.5 \times 10^{-9}$ MeV deduced in
\cite{Chromik:1997}.

A more advanced than in \cite{Grigorenko:2000b} structure model of $^{17}$Ne was
developed in \cite{Grigorenko:2003}. This was taking into account core spin and
provided accurate treatment of the excitation spectra in $^{16}$F subsystem of
$^{17}$Ne. In the improved model, the width estimate for the $3/2^-$ state decay
was reduced to $\Gamma_{2p}=4.1 \times 10^{-16}$ MeV. The results
\cite{Grigorenko:2000b,Grigorenko:2003} were obtained with relatively limited
basis sizes computationally feasible at that time.

Shortly later the work \cite{Garrido:2004b} criticized the results of
\cite{Grigorenko:2003} and also provided very different $\Gamma_{2p}=3.6 \times
10^{-12}$ MeV. This was unexpectedly large width value, larger than the one
provided by the diproton model (as we mentioned, the latter is known to give the
strict upper limit for the $2p$ decay width).

The strong disagreement between $\Gamma_{2p}$ in Refs.\ \cite{Grigorenko:2003}
and \cite{Garrido:2004b} inspired us to perform considerable theoretical
developments, including large basis three-body calculations and construction of
``exact'' semianalytical models which were free of basis convergence issues
\cite{Grigorenko:2007,Grigorenko:2007a}. The revised $2p$ width calculations
allowed to confine the possible $^{17}$Ne $3/2^-$ state width in the limits
$\Gamma_{2p}=(5-8) \times 10^{-15}$ MeV. There was no chance to close the gap
with the results of \cite{Garrido:2004b}. Consequently, the results of Ref.\
\cite{Garrido:2004b} were revised in Ref.\ \cite{Garrido:2008} claiming the
important effect of basis convergence. The revised width value $\Gamma_{2p}=1.7
\times 10^{-14}$ MeV was concluded to be consistent with the results of
\cite{Grigorenko:2007}. So, it may seem that the controversy
\cite{Grigorenko:2003,Garrido:2004b,Grigorenko:2007,Garrido:2008} about the
$^{17}$Ne $3/2^-$ state width was resolved.

In recent experimental work \cite{Sharov:2017}, the significantly improved limit
$\Gamma_{2p} / \Gamma_{\gamma} <1.6 \times 10^{-4}$ was established. Using the
$^{17}$Ne $3/2^-$ state gamma width from \cite{Chromik:1997} the limit
$\Gamma_{2p} <8.8 \times 10^{-13}$ MeV can be found. This experimental advance
inspired us to revisit the issue of the $^{17}$Ne $3/2^-$ state width. We
found important inconsistencies in the quasiclassical three-body widths
treatment in \cite{Garrido:2004b,Garrido:2008}. Moreover, in the course of this
activity, we arrived at a conclusion about the poor applicability of the
quasiclassical approach to the $^{17}$Ne $3/2^-$ state decay. This sheds doubts on
the applicability of such an approach to three-body decays in general.


\section{Width definitions}


The most widespread width definitions are connected with the determination of
elastic scattering phase shifts $\delta_l(E)$. Then the resonance width can be
defined either as FWHM for the resonance peak in the elastic cross section
\begin{equation}
\sigma_{l}(E) \sim (\pi/k^2) \,\sin^2 \delta_l(E) \,,
\label{eq:sigma}
\end{equation}
or, based on the R-matrix expression for phase shift on proximity of resonance
\begin{equation}
\delta_l(E) = \arctan \left[ \frac{\Gamma}{2(E_r-E)} \right] \quad \rightarrow
\quad \left. \frac{d \delta_l(E)}{dE} \right|_{E_r} = \frac{2}{\Gamma}\, .
\label{eq:delta}
\end{equation}
Both methods are very inconvenient for small decay widths $\Gamma \ll E_r$,
where a search for the resonance position and ``energy scan'' in its proximity
becomes computationally a bit not straightforward.

In this case, the quasiclassical Gamow formula may be applied
\begin{equation}
\Gamma = \nu \exp \left[ 2i \int \nolimits ^{r_3}_{r_2} p(r) \, d r \, \right]
, \,\; \nu = \left[ 2M \int \nolimits ^{r_2}_{r_1} \frac{dr}{p(r)} \right]^{-1}
, \,\; p(r)=\sqrt{2M[E_T-V(r)]}\,.
\label{eq:qc}
\end{equation}
Here $r_1$, $r_2$ are inner, $r_3$ is outer classical turning points of the
potential $V(r)$ and $M$ is the reduced mass for the channel. The preexponent
$\nu$ with a dimension of energy is typically evaluated as an ``assault frequency''
of classical motion in the potential well $\{r_1, r_2 \}$. For the $2p$ width
calculations, both the validity of the three-body problem reduction to a single
channel formalism and the applicability of the QC approximation for barriers of
specific shape can be questioned.

There exists a somewhat more complicated approach of integral (sometimes called
Kadmensky-type) formulas for the decay width determination
\cite{Harada:1968,Kadmenskii:1971,Grigorenko:2007}, which allows solving the
Schr\"odinger equation only for one selected (resonance) energy. In that sense,
the IF method is an analogue of QC approach and can be used to cross-check the
QC results. The width value in IF method is defined as
\begin{equation}
\Gamma=\frac{4M}{k\,\cos^2(\delta_l)} \left| \int \nolimits_0^{r_{\mbox{\tiny
in}}} \varphi^*_l(kr) \, (V-\bar{V}) \,\tilde{\psi}_l(kr) \, dr \right|^2\,,
\label{eq:kadm}
\end{equation}
where \emph{quasistationary} WF $\tilde{\psi}_l(kr)$ is normalized in the
``internal region'' $r_{\mbox{\scriptsize in}}$ and obtained by solving the
Schr\"odinger equation with potential $V$ with the quasistationary boundary
condition
\begin{equation}
\tilde{\psi}_l(kr_{\mbox{\scriptsize in}}) \sim G_l(kr_{\mbox{\scriptsize in}})
\,, \quad
\int \nolimits^{r_{\mbox{\tiny in}}}_0 |\tilde{\psi}_l(kr)|^2 \, dr \equiv 1 \,.
\label{eq:qss}
\end{equation}
In general case, the $G_l$ is irregular at the origin Coulomb WF. The
\emph{auxiliary} scattering WF $\varphi_l(kr)$ is obtained with potential
$\bar{V}$ and normalized by diagonalizing S-matrix and providing phase shifts
$\delta_l$. This formula has an especially simple form in the case $\bar{V}
\equiv
V_{\mbox{\scriptsize coul}}=Z_1Z_2\alpha/r$
\begin{equation}
\Gamma=\frac{4M}{k} \left| \int \nolimits_0^{r_{\mbox{\tiny in}}} F_l(kr) \,
(V-V_{\mbox{\scriptsize coul}}) \, \tilde{\psi}_l(kr) \, dr \right|^2\,,
\label{eq:kadm-2}
\end{equation}
where $F_l$ is regular at the origin Coulomb WF. It is clear that the radial
convergence of the integral in Eq.\ (\ref{eq:kadm-2}) is provided when the value
of $r_{\mbox{\scriptsize in}}$ is selected outside the nuclear interaction
region, where  $V-V_{\mbox{\scriptsize coul}} \equiv 0$. In this case, some
uncertainty remains in the Eq.\ (\ref{eq:kadm-2}), which is connected with an
uncertainty of normalization of $\tilde{\psi}_l$ in the Eq.\ (\ref{eq:qss}).
This uncertainty is quite sizable for the case of small barriers. It is possible
to show, see \cite{Grigorenko:2007}, that in such a case $r_{\mbox{\scriptsize in}}
\sim r_3$ should be selected for the best match between Eq.\ (\ref{eq:sigma})
and Eq.\ (\ref{eq:kadm-2}) results.


\subsection{Reliability of the Gamow formula}


Let us consider first applicability of the QC formula (\ref{eq:qc}) in different
conditions.

\begin{figure}[tb]
\begin{center}
\includegraphics[width=0.7\textwidth]{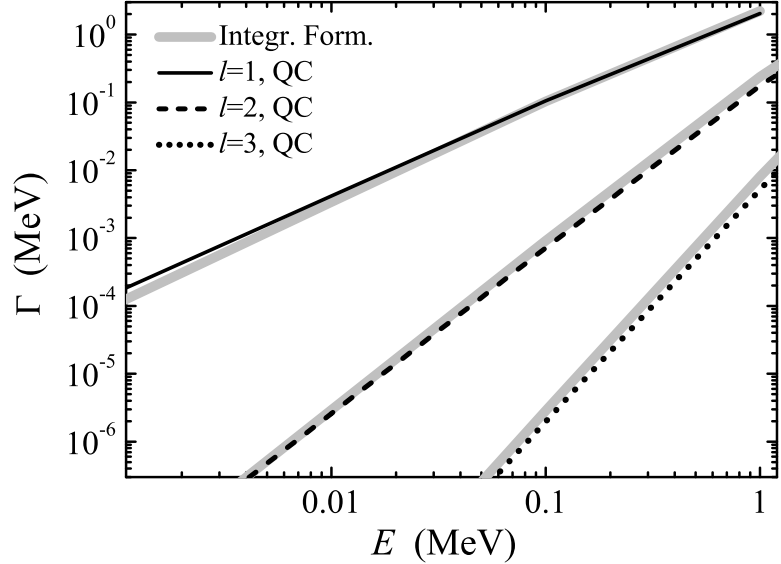}
\end{center}
\caption{Decay width calculations for dineutron+dineutron system with radius
$r_0= 4 $ fm and diffuseness $a=0.001$ fm for different angular momenta by
integral formula and quasiclassical approximation.}
\label{fig:4n}
\end{figure}

Fig.\ \ref{fig:4n} shows the application of Eq.\ (\ref{eq:qc}) to the system
``dineutron''+``dineutron'' (no Coulomb interaction, the reduced mass is just
equal to the nucleon mass). Integral formula results are provided for
cross-checking with standard potential formalism. It was also checked that for
relatively large widths (e.g.\ corresponding to the decay energies $E>10$ eV)
the IF provides exactly the same results as standard potential scattering
calculations (\ref{eq:sigma}).

\begin{figure}[tb]
\begin{center}
\includegraphics[width=0.7\textwidth]{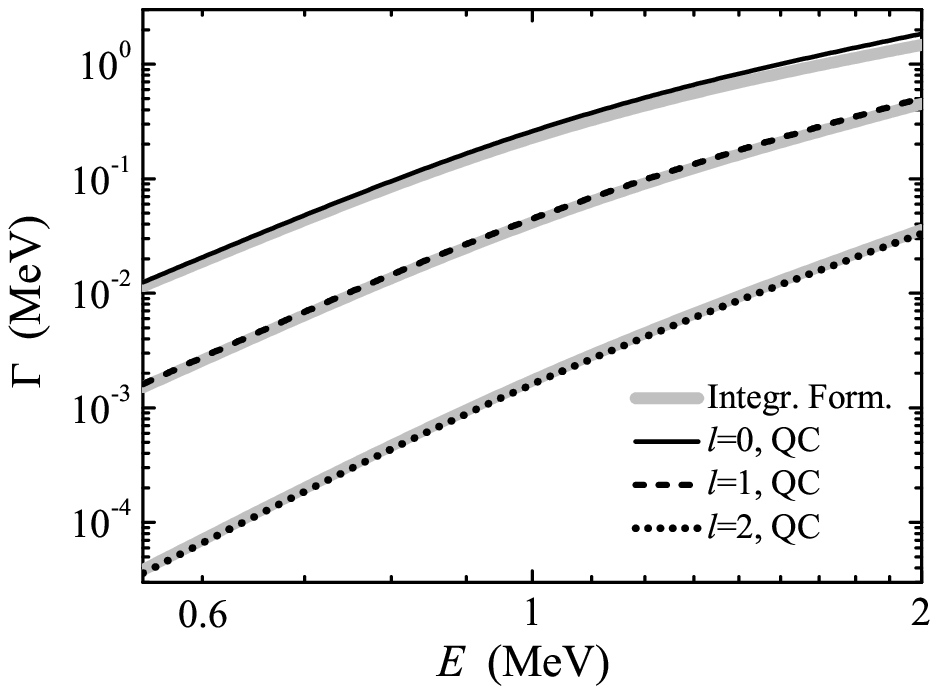}
\end{center}
\caption{Decay width  calculations for $^{15}\mbox{F} \rightarrow \,
{^{14}\mbox{O}}+p$ system with $r_0= 2.96 $ fm and diffuseness $a=0.001$ fm for
different angular momenta by integral formula and quasiclassical approximation.}
\label{fig:15f-a}
\end{figure}

\begin{figure}[tb]
\begin{center}
\includegraphics[width=0.7\textwidth]{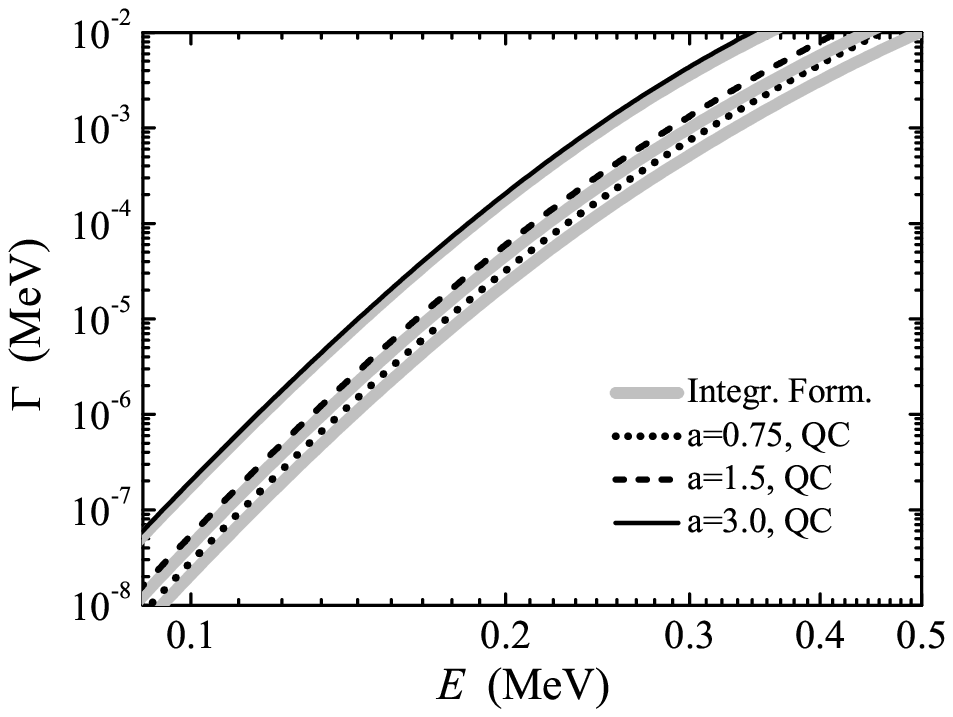}
\end{center}
\caption{Decay width calculations for $^{15}\mbox{F}\rightarrow \,
{^{14}\mbox{O}}+p$ system with $r_0= 2.96 $ fm and angular momentum $l=0$ for
different diffusenesses $a$ by integral formula and quasiclassical
approximation.}
\label{fig:15f-l}
\end{figure}

The results of two tests with Coulomb interaction are shown in  Figs.\
\ref{fig:15f-a} and \ref{fig:15f-l}. Fig.\ \ref{fig:15f-a} shows the
calculations with different angular momenta. Fig.\ \ref{fig:15f-l} shows the
calculations with fixed angular momentum, but for different diffusenesses. Here
we test also the cases of diffusenesses much larger than those typical for
two-body nuclear potentials, e.g.\ $a \lesssim 1$ fm.

We can see that the QC formula is quite precise by itself (from a few percents
to few tens of percent). It has a trend to become more precise for larger barriers
and lower (tending to zero) decay energies. There is also no problem to operate
it for potentials with large diffuseness (in three-body case there is no well
defined ``nuclear radius'' for effective potential and large diffesenesses may
take place).

\section{Three-body problem reduction}



\subsection{Hyperspherical harmonics method}


The application of the hyperspherical harmonics method to three-body system
provides a set of coupled differential equations in $\rho$ variable, see e.g.\
\cite{Grigorenko:2009c} for more details
\begin{equation}
\left[ \frac{d^{2}}{d\rho^{2}} - \frac{\mathcal{L(L}+1)}{\rho^{2}}
 + 2M \{E-V_{K\gamma,K\gamma}(\rho) - V_{\mbox{\scriptsize 3b}} \}
 \right]\chi_{K\gamma}(\rho)
= \sum_{K'\gamma\,'} 2M V_{K\gamma,K'\gamma\,'} (\rho) \chi_{K'\gamma\,'}(\rho)
\,,
\label{eq:hh}
\end{equation}
where $E_T$ is the energy relative to the three-body breakup threshold, $M$ is a
``scaling'' average nucleon mass, and $V_{\mbox{\scriptsize 3b}}$ is short-range
phenomenological three-body potential used to fine-tune the decay energy.
Effective angular momentum $\mathcal{L}$ in HH equations is expressed via the
principal hyperspherical quantum number $K$ as
\begin{equation}
\mathcal{L}=K+3/2 \,,
\label{eq:l-eff}
\end{equation}
so, in contrast with a two-body case, the centrifugal barrier in the three-body
case is never equal to zero. The hyperspherical potentials are matrix elements
of the pairwise intercluster potentials over hyperspherical harmonics
$\mathcal{J}_{K\gamma}$
\begin{equation}
V_{K\gamma,K'\gamma\,'}(\rho) =
\int d\Omega_\rho \quad
\mathcal{J}_{K\gamma}^\dag(\Omega_\rho)
\left[ \sum \nolimits_{i\,>j} \hat{V} (\textbf{r}_{ij})\right]
\mathcal{J}_{K'\gamma\,'}(\Omega_\rho) \,,
\label{eq:hh-pot}
\end{equation}
where $\Omega_\rho$ is the 5-dimensional ``hyperangle'', which together with
hyperradius $\rho$ provides the complete description of the internal degrees of
freedom for three-body systems.

\begin{figure}
\begin{center}
\includegraphics[width=0.8\textwidth]{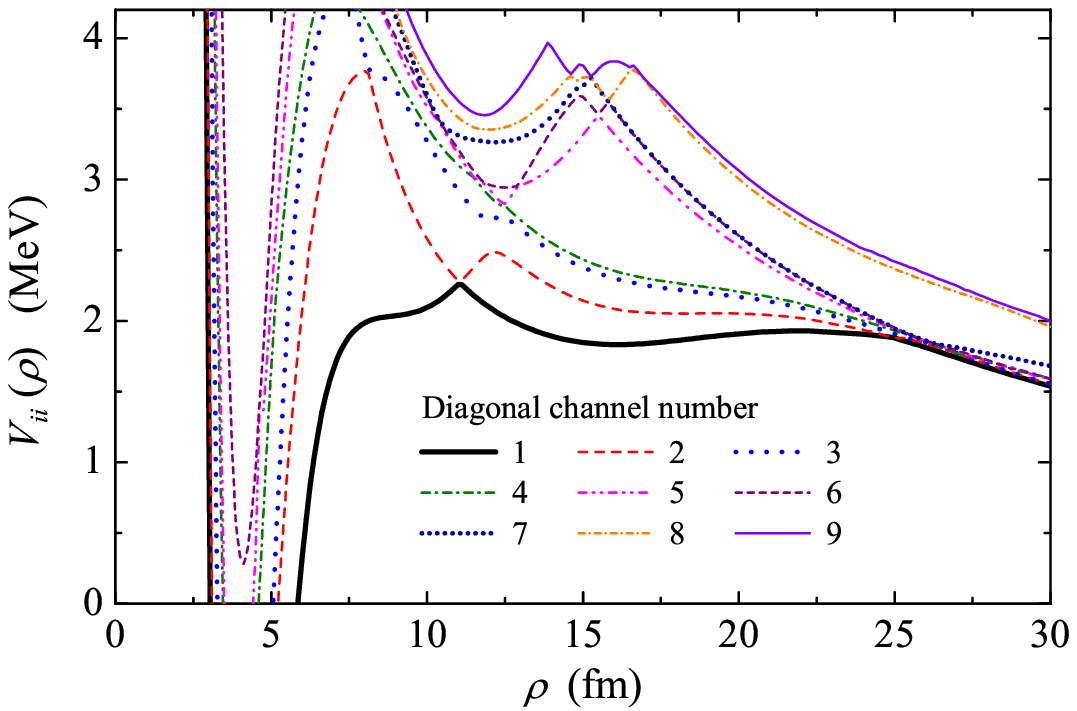}
\end{center}
\caption{Several lowest terms of the diagonalized potential matrix Eq.\
(\ref{eq:pot-diag}) in the HH method for $K_{\max}=20$. It is assumed QC
tunneling occurs along the lowest diagonal term (black solid curve).}
\label{fig:diagonal}
\end{figure}

The easiest way for a transition to single channel representation suitable for
quasiclassical treatment is a potential diagonalization
\begin{equation}
\tilde{V}=O^{T}VO \quad, \qquad
\tilde{V}_{ij}=\delta_{ij} V_{ii}\,,
\label{eq:pot-diag}
\end{equation}
provided by the orthogonal matrix $O$. Some lowest terms of the diagonalized
potential matrix for HH potentials from \cite{Grigorenko:2003,Grigorenko:2007}
are shown in Fig.\ \ref{fig:diagonal}. The diagonal terms of the potential
matrix are intersecting and the effective potential is taken be assuming that
quasiclassical motion is taking place all the time along the lowest-energy
branch of the potential matrix.

\subsection{Hyperspherical adiabatic expansion method}


Within the hyperspherical adiabatic expansion (HAE) method the following
equations are solved (e.g., \cite{Garrido:2004b,Garrido:2008})
\begin{equation}
\left[
 -\frac{d^{2}}{d\rho^{2}}
 + 2M \left[ V_{\mbox{\scriptsize 3b}}(\rho) - E_T \right]
 + \frac{1}{\rho^{2}}\left\{ \lambda_n(\rho)+\frac{15}{4} \right \} \right]
 f_n(\rho) + \sum \nolimits_{n'} \left(-2P_{nn'}\frac{d}{d\rho}
 -Q_{nn'}\right)f_{n'}(\rho) = 0 \,.
\label{eq:hyp-ad}
\end{equation}
These equations contain the effective adiabatic potentials, which take the form
\begin{equation}
V_{\mbox{\scriptsize eff}}(\rho)
=\frac{\lambda_n(\rho)+15/4}{ 2M \,\rho^{2}}
+V_{\mbox{\scriptsize 3b}}(\rho) \,.
\label{eq:v-eff-ad}
\end{equation}
The adiabatic terms $\lambda_n(\rho)$ in this approach have complicated radial
behavior: they may intersect. For width calculations the lowest-energy branch of
the adiabatic potential is taken.

Although this procedure was used in numerous works
\cite{Garrido:2004b,Garrido:2005,Garrido:2005a,Garrido:2008,Garrido:2010,%
Garrido:2011,Hove:2016,Hove:2017}, it has never been justified theoretically 
and was just accepted as a reasonable approach.


\section{Three-body width of the $^{17}$Ne $3/2^-$ state}

\begin{figure}[t]
\begin{center}
\includegraphics[width=0.8\textwidth]{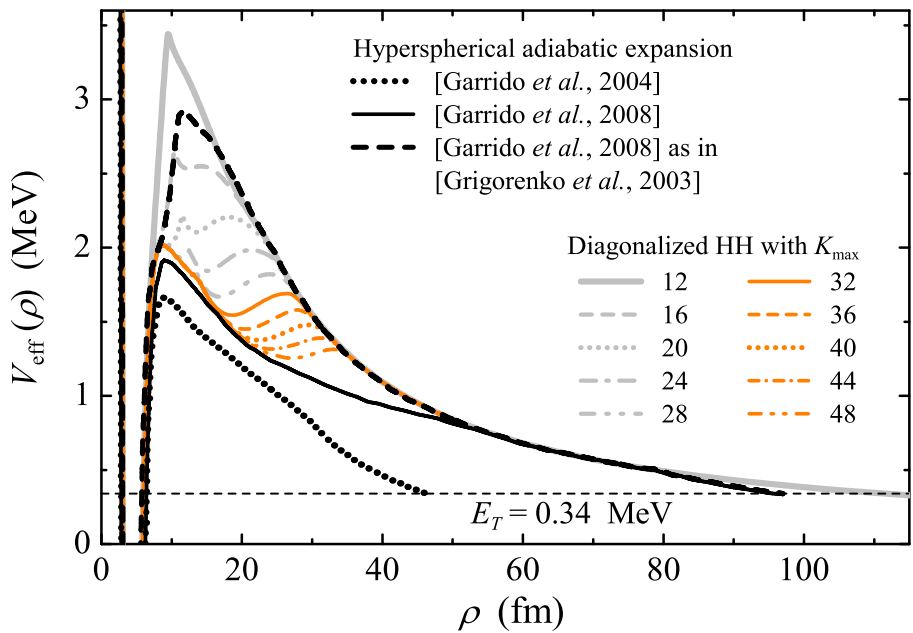}
\end{center}
\caption{Effective single channel potentials $V_{\mbox{\scriptsize eff}}$ for
quasi-classical calculations of two-proton width of the $^{17}$Ne $3/2^-$ state.
The two-proton decay energy $E_T$ is indicated by horizontal dashed line. Our
hyperspherical harmonics potentials (gray and orange curves) are obtained by
trivial diagonalization of the HH potential matrix from \cite{Grigorenko:2007}.
The dotted, solid and dashed black lines are from
\cite{Garrido:2004b,Garrido:2008}. They were obtained for the basis sizes
$K_{\max} = 70, \, l_x,l_y \leq 2 \, $,
$K_{\max} = 70, \, l_x,l_y \leq 9 \, $, and
$K_{\max} = 20, \, l_x,l_y \leq 10$, correspondingly.}
\label{fig:pot-garrido}
\end{figure}

The paper \cite{Garrido:2008} is dedicated to finding ``necessary conditions for
accurate computations of three-body partial decay widths''. The Fig.\
\ref{fig:pot-garrido} shows the effective potential curves from papers
\cite{Garrido:2004b,Garrido:2008} (see Fig.\ 4 of the latter). The black dotted
curve corresponds to the result of \cite{Garrido:2004b} which was claimed to be
``non-converged'' in \cite{Garrido:2008} (basis size $K_{\max} = 70, \, l_x,l_y
\leq 2 $). The solid black curve corresponds to the ``accurate converged''
result from \cite{Garrido:2008}  (basis size $K_{\max} = 70, \, l_x,l_y \leq 9
$). The dashed black curve was to imitate calculation conditions of
\cite{Grigorenko:2003} with $K_{\max} = 20, \, l_x,l_y \leq 10 $. We repeated
the QC calculations with $V_{\mbox{\scriptsize eff}}$ of the Fig.\
\ref{fig:pot-garrido}, see Table \ref{tab:qc}. For the decay of the $^{17}$Ne
$3/2^-$ state we have found that we cannot reconcile the width values quoted in
Ref.\ \cite{Garrido:2008} with the potential curves provided in this work.

\begin{table}
\caption{The $^{17}$Ne $3/2^-$ state widths (in MeV) obtained by Eqs.\
(\ref{eq:kadm-2}) and (\ref{eq:qc}) for hyperspherical adiabatic effective
potentials $V_{\mbox{\scriptsize eff}}$ from \cite{Garrido:2004b,Garrido:2008}.
Column 1 corresponds to the ``non-converged'' result of \cite{Garrido:2004b}
(dotted black curve in Fig.\ \ref{fig:pot-garrido}),
column 2 to the ``accurate converged'' result from \cite{Garrido:2008} (solid
black curve in Fig.\ \ref{fig:pot-garrido}), and
column 3 shows the result from \cite{Garrido:2008} which aims to imitate the
calculations of Ref.\ \cite{Grigorenko:2003} (dashed black curve in Fig.\
\ref{fig:pot-garrido}).}
\label{tab:qc}
\begin{center}
\begin{tabular}{cccc}
\hline
Case: &
$K_{\max} = 70, \, l_x,l_y \leq 2 $  &
$\quad K_{\max} = 70, \, l_x,l_y \leq 9 \quad $  &
$K_{\max} = 20, \, l_x,l_y \leq 10$ \\
\hline
Ref.\ \cite{Garrido:2004b,Garrido:2008} & $3.6 \times 10^{-12}$ & $1.7 \times
10^{-14}$ & $5.4 \times 10^{-16}$ \\
IF, Eq.\ (\ref{eq:kadm-2}) & $5.1 \times 10^{-6}$ & $5.7 \times 10^{-12}$ &
$7.7 \times 10^{-14}$ \\
QC, Eq.\ (\ref{eq:qc}) & $5.6 \times 10^{-6}$ & $6.3 \times 10^{-12}$ &  $7.2
\times 10^{-14}$ \\
\hline
\end{tabular}
\end{center}
\end{table}

In more details the procedure was like follows. At first, we performed the
calculations with the integral formula for width,
see Eq.\ (\ref{eq:kadm-2}), appropriately modified for the three-body case.
Effective potentials from Ref.\ \cite{Garrido:2008} (see Fig.\ 4 therein) were
scanned and interpolated. Since the $V_{\mbox{\scriptsize eff}}$ behavior inside
the potential well ($E_T<0$) cannot be found in Ref.\ \cite{Garrido:2008}, we
used the short-range potential $V_{\mbox{\scriptsize 3b}}$ with the Woods-Saxon
formfactor to reproduce the $Q_{2p}=0.34$ MeV in the integral formula formalism.
The potentials $V_{\mbox{\scriptsize 3b}}$ were also selected in such a way,
that for $E_T>0$
they do not affect the $V_{\mbox{\scriptsize eff}}$ behavior in the barrier
region. The long-range behavior of the $V_{\mbox{\scriptsize eff}}$ was fitted
to the expected asymptotic form of the three-body potential in the systems with
Coulomb interaction
\begin{equation}
V_{\mbox{\scriptsize eff}}(\rho) =
\frac{\mathcal{L}(\mathcal{L}+1)}{2M\rho^{2}}+ \frac{Z_{\mbox{\scriptsize eff}}
\alpha }{\rho}
+V_{\mbox{\scriptsize 3b}}(\rho)\,.
\label{eq:v-ass}
\end{equation}
The $\mathcal{L}=7/2$ is taken as we know that only the penetration through the
three-body centrifugal barrier defined by the lowest possible hyperspherical
excitation $K=2$ is important on long-range asymptotics. The values
$\mathcal{L}=7/2$ and $Z_{\mbox{\scriptsize eff}}$ are used to define the
Coulomb WFs used in the
IF calculations by Eqs.\ (\ref{eq:qss}) and (\ref{eq:kadm-2}). For consistency,
the QC calculations were performed with the same $V_{\mbox{\scriptsize 3b}}$
which were fitted to the correct $Q_{2p}$ value in IF calculations. The results
of QC calculations are typically within $10 \%$ around the IF values. So, QC
approximation is quite precise and cannot be a source of the problems here.

Can it be the effective potential curves are provided in Ref.\
\cite{Garrido:2008} somehow in a wrong way or we interpret them incorrectly? We
have performed our own HH calculations with three-body potentials from Ref.\
\cite{Grigorenko:2003,Grigorenko:2007} using the diagonalization procedure Eq.\
(\ref{eq:pot-diag}), see gray and orange curves in Fig.\ \ref{fig:pot-garrido}.
The same procedure was used for the determination of  $Z_{\mbox{\scriptsize
eff}}$, as
described above in Eq.\ (\ref{eq:v-ass}). Despite the fact that our three-body
potentials are based on somewhat different two-body interactions, there are
large overlap regions for $V_{\mbox{\scriptsize eff}}$ produced by HH and HAE.
There is a very good overlap in the region of the Coulomb interaction
dominance, where different methods should be providing close results. In the
particular case of Fig.\ \ref{fig:pot-garrido}, exact overlap can be seen for
$50< \rho < 80$ fm. Evidently, we correctly interpret the effective potentials
provided in Ref.\ \cite{Garrido:2008}.

The potentials obtained by a trivial single-channel reduction of the
hyperspherical potentials of Ref.\ \cite{Grigorenko:2007} (just diagonalization)
have an analogous basis convergence trend with potential derived in
\cite{Garrido:2008}. It can be seen that $K_{\max}=20$ calculations in HAE
(black dashed curve in Fig.\ \ref{fig:pot-garrido}) and HH (gray dotted curve)
do not
correspond well to each other in contrast with expectations of
\cite{Garrido:2008}. However, with the basis increase, the HH results are
beginning
to follow the HAE results in a larger and larger interval of $\rho$ values. It
is
reasonable to assume that they finally converge to the same profile, which is
quite close to the converged HAE result (black solid curve in Fig.\
\ref{fig:pot-garrido}).

So, the single-channel effective potentials obtained in HAE and HH are
consistent and should provide consistent QC and IF results. However, the QC
results for $\Gamma_{2p}$ obtained with effective HH potentials are not
consistent with the results of dynamical three-body calculations, which were
accurately validated in Ref.\ \cite{Grigorenko:2007}. This comparison for the
different basis sizes is provided in Table \ref{tab:comp-3b}. The disagreement
is modest for small $K_{max}$ values, but for asymptotically large $K_{max}$ the
difference exceeds two orders of the magnitude.  So, we find that the reduction
of three-body problem to one-channel approximation leads to the significant
width overestimation, compared to fully dynamical three-body calculations.

\begin{table}
\caption{Widths of the $^{17}$Ne $3/2^-$ state in $10^{-15}$ MeV units as a
function of the hyperspherical basis size $K_{\max}$. IF and QC calculations
with diagonalized HH potentials from \cite{Grigorenko:2007}, three-body
calculations of Ref.\ \cite{Grigorenko:2007} (see Fig.\ 15), and their ratio.
Column ``Asympt.'' contains values obtained by exponential extrapolation to the
infinite basis.}
\label{tab:comp-3b}
\begin{center}
\begin{tabular}{ccccccccc}
\hline
$K_{\max}$ & $\quad 12 \quad$ & $\quad 16 \quad$ & $\quad 20\quad$ & $\quad
24\quad$ & $\quad 32\quad$ & $\quad 40\quad$ & $\quad 48\quad$ & $\;$ Asympt.
$\;$ \\
\hline
IF, Eq.\ (\ref{eq:kadm-2}) & 14.1 & 27.0 & 49.9 & 81.7 & 171 & 285 & 420 & 1100
\\
QC, Eq.\ (\ref{eq:qc})     & 9.83 & 21.8 & 42.2 & 71.0 & 151 & 251 & 369 & 971
\\
3-body, Ref.\ \cite{Grigorenko:2007}  & 0.91 & 1.32 & 1.70 & 2.15 & 3.04 & 3.86
& 4.55 & 6.90
\\
Ratio QC/3-body & 10.8 & 16.5 & 24.9 & 33.0 & 49.6 & 65.0 & 81.0 & 141 \\
\hline
\end{tabular}
\end{center}
\end{table}

We can draw several conclusions of different nature from our calculations here:

\noindent (i) The effective potential provided in Ref.\ \cite{Garrido:2004b}
(see the black dotted curve in Fig.\ \ref{fig:pot-garrido}) was a pure mistake.
For
example, the long-range asymptotic behavior of this potential cannot be
correct. Thus there is no way to obtain it as a reduction of any three-body
potential whatever is the convergence. In addition, the QC width calculation by
itself for this potential was also erroneous ($\sim 6$ orders of the magnitude
away) in \cite{Garrido:2004b}.

\noindent (ii) The effective potential provided in Ref.\ \cite{Garrido:2008}
(see the black solid curve in Fig.\ \ref{fig:pot-garrido}) looks reasonable.
However, the QC width calculations for this potential were also erroneous ($\sim
2$ orders of the magnitude away).

\noindent (iii) The results for the decay widths analogous to the results of
\cite{Garrido:2004b,Garrido:2008} obtained in HAE can be obtained in HH method
just without any dynamical calculation by trivial diagonalization of the
potential matrix.

\noindent (iv) The QC and IF widths for single-channel reduction of the
three-body problem produce much larger (up to more than two orders of the
magnitude) width values than the dynamical HH calculations
\cite{Grigorenko:2007}. This is true in a broad range of basis selections.

\noindent (v) The width value $\Gamma_{2p}=6.3 \times 10^{-12}$ MeV is
recalculated by us from the best-converged effective potential from
\cite{Garrido:2008}. This value exceeds the recent experimental limit
$\Gamma_{2p} < 8.8 \times 10^{-13}$ MeV from \cite{Sharov:2017} by an order of
the magnitude. This also adds confidence in the erroneous character of this
result.


\section{Conclusion}


The standard quasiclassical approximation is easy to formally generalize for the
true three-body decays, the decays in which two protons are emitted
simultaneously. The motion of such a system can be considered in a certain
approximation as a single-channel motion in the hyperradius $\rho$ value. In
this work, we have explored the application of quasiclassical approximation to
to the true three-body decays.

As a first step, we have systematically compared the ``ordinary'' two-body width
calculations in quasiclassical approximation with potential model calculations
by means of an integral formula. These approaches are found to be highly
consistent
(within $10 \%$) for different combinations of angular and Coulomb barriers and
nuclear potential diffuseness. So, for the effective hyperspherical barrier with
large effective angular momentum, charge, and diffuseness the quasiclassical
approximation by itself is not expected to be an obstacle.

Specifically for the decay of the $^{17}$Ne $3/2^-$ state we have found problems
of two kinds, specific for Refs.\ \cite{Garrido:2004b,Garrido:2008} and generic
for quasiclassical approximation:

\noindent (i) We can not reconcile the width values quoted in Ref.\
\cite{Garrido:2008} with the potential curves provided in this work. In general
the paper \cite{Garrido:2008} is dedicated to finding ``necessary conditions for
accurate computations of three-body partial decay widths'', so its results are
expected to be specifically accurate. The potential provided as a final result
of these studies for the $^{17}$Ne $3/2^-$ state decay gives the width value
which (according to our calculations) exceeds the recent experimental limit for
about an order of the magnitude.

\noindent (ii) The potentials obtained by a single-channel reduction of the
hyperspherical potentials of Ref.\ \cite{Grigorenko:2007} look reasonably
consistent with potential derived in \cite{Garrido:2008}, despite the method is
quite different. However, in the case of the HH method the cross-check with
complete three-body calculations is available. The quasiclassical results
obtained with HH method effective potentials are found to be more than an order
of
the magnitude larger than the results of the complete three-body calculations
\cite{Grigorenko:2007} for each considered basis size.

Finally, we conclude that quasiclassical approximation for single channel is
quite precise, but the reduction to one-channel approximation leads to
significant
overestimation of the three-body width. We think that in the view of our results
all the three-body width calculations in the papers
\cite{Garrido:2004b,Garrido:2005,Garrido:2005a,Garrido:2008,Garrido:2010,%
Garrido:2011,Hove:2016,Hove:2017} should be questioned and applicability of the
quasiclassical formalism in this case reexamined in general.

\textit{Acknowledgements.} O.M.S.\ and L.V.G.\ were partly supported by the
Russian Science Foundation grant No.\ 17-12-01367.


\bibliographystyle{elsarticle-num}
\bibliography{d:/latex/all}


\end{document}